# The Self-Organization of Meaning and the Reflexive Communication of Information

*Social Science Information* (in press)

Loet Leydesdorff,*[a]  Alexander M. Petersen,[b] and Inga Ivanova [c]

**Abstract**

Following a suggestion of Warren Weaver, we extend the Shannon model of communication piecemeal into a complex systems model in which communication is differentiated both vertically and horizontally. This model enables us to bridge the divide between Niklas Luhmann's theory of the self-organization of meaning in communications and empirical research using information theory. First, we distinguish between communication relations and correlations among patterns of relations. The correlations span a vector space in which relations are positioned and can be provided with meaning. Second, positions provide reflexive perspectives. Whereas the different meanings are integrated locally, each instantiation opens global perspectives—"horizons of meaning"—along eigenvectors of the communication matrix. These next-order codifications of meaning can be expected to generate mutual redundancies when interacting in instantiations. Increases in redundancy indicate new options and can be measured as local reduction of prevailing uncertainty (in bits). The systemic generation of new options can be considered as a hallmark of the knowledge-based economy.

[a] * corresponding author; Amsterdam School of Communication Research (ASCoR), University of Amsterdam PO Box 15793, 1001 NG Amsterdam, The Netherlands; loet@leydesdorff.net

[b] Laboratory for the Analysis of Complex Economic Systems, IMT Institute for Advanced Studies Lucca, 55100 Lucca, Italy; petersen.xander@gmail.com

[c] Institute for Statistical Studies and Economics of Knowledge, National Research University Higher School of Economics (NRU HSE), 20 Myasnitskaya St., Moscow, 101000, Russia; and School of Economics and Management, Far Eastern Federal University, 8, Sukhanova St., Vladivostok 690990, Russia; inga.iva@mail.ru

**Introduction**

In his contribution to Shannon & Weaver's (1949) *The Mathematical Theory of Communication*, Warren Weaver stated (at p. 27) that "[t]he concept of information developed in this theory at first seems disappointing and bizarre—disappointing because it has nothing to do with meaning …" However, the author added that Shannon's analysis "has so penetratingly cleared the air that one is now, perhaps for the first time, ready for a real theory of meaning." But how can one relate a theory of meaning to Shannon's information theory (cf. Bar-Hillel & Carnap, 1953)? More recently, Niklas Luhmann ([1984] 1995) argued that meaning ("*Sinn*") self-organizes in terms of communications among human beings. From this perspective meaning is generated in interactions among communications. As Luhmann (1996, at p. 261) formulated: "My argument is: it is not human beings who can communicate, rather, only communication can communicate." However, Luhmann's theory has remained far from operationalization and measurement.

Following Bateson's (1972, at p. 315) alternative definition of information as "a difference which makes a difference" (cf. MacKay, 1969), Luhmann (1984, pp. 102 ff.; 1995, pp. 67f.) considered information as implying a selection: a difference can only make a difference for a system of reference that selects this difference from among other possible differences. Others have also defined information with reference to a receiving system (e.g., an observer). Varela (1979, p. 266) even argued that since the word "information" is derived from "in-formare," the semantics call for the specification of a system of reference to be informed. "Information," however, is then considered a substantive concept that varies with the system of reference instead of a formal measure of the uncertainty prevailing in a distribution. Kauffman *et al.* (2008, at p. 28), for example, defined information as "natural selection assembling the very constraints on the release of energy that then constitutes work and the propagation of organization." In summary, using this alternative definition of information, the meaning of "information" becomes dependent on the theoretical context. Using the same word ("information") for different concepts has led to considerable confusion in the literature.

In an assessment of this confusion, Hayles (1990, pp. 59f.) compared the discussion with asking whether a glass is half empty or half full. As she noted, confusion can be avoided by using the words "uncertainty" or "probabilistic entropy" when Shannon-type information is meant. In our opinion, the advantage of measuring uncertainty—and redundancies, as we shall argue—in bits of information cannot be underestimated, since the operationalization and the measurement provide avenues to hypothesis testing and thus control of the theorizing (Theil, 1972). Note that uncertainty cannot be specified in terms of Bateson's definition, but his "a difference which makes a difference" can be operationalized and measured in terms of (potentially negative) bits of information (Brillouin, 1962; von Foerster, 1960).

Can Luhmann's theory about interacting communications and the self-organization of meaning also be made compatible with Shannon's information theory? Is it possible to specify how information and meaning are related? In this study, we aim to contribute to bridging this gap between a focus on meaning versus uncertainty processing by decomposing the problem using Herbert Simon's (1973a) model of complex systems that are both vertically and horizontally differentiated, as follows:

1. In the vertical dimension, we follow Luhmann's (e.g., 1975; 2000) distinction between (*i*) interactions among communications providing variation, (*ii*) the organization of meanings in historical instantiations, and (*iii*) the self-organization of reflexive meaning generating a next level of "horizons of meaning" as global systems of reference (Husserl, 1929; cf. Luhmann, 1995b). Despite this inspiration from Luhmann, however, we also deviate from his framework, and argue that the construction in terms of layers is bottom-up from the (probabilistic) informational level; but the emerging system's levels can be expected to take over control in terms of codified intentionalities and expectations. In other words, the operation in layers can also be described in terms of variation and selection mechanisms.

   For example, a scholarly communication (e.g., a manuscript) can be expected to contain a knowledge claim. Knowledge claims provide in this case the historical variation. When the manuscript is submitted, an editorial process is instantiated in which referee comments, editorial judgments, etc., are combined. The referees, however, are expected to judge the manuscript in terms according with the standards of the field invoking codes of the communication that can be expected to control the process. The instantiation requires a reflexive reorganization of the codes.

   Unlike Simon's model (of complex and artificial systems), these next-order constructs—codes of communication—can operate with frequencies much higher than the underlying practices. One can consider the constructs from two sides: the first being the constructing agency (in history), and the second being the resulting constructs themselves, which can only be entertained reflexively (Giddens, 1979).[1] Because of this reflexive status, the processing (e.g., sharing of meaning) is volatile and cannot be observed directly; but the operations can be specified. On the basis of this specification, one is able to identify and measure the "footprints" of the self-organizing dynamics in historical instantiations.

In other words, intentions and intentional systems cannot be found as observables in *res extensa*—or "matter"—but can only be hypothesized reflexively in *res cogitans*—"thought" (Husserl, 1929; Luhmann, 1990). Entertaining such hypotheses can enrich our expectations by providing frames for inferences about observations. The consequent possibility of an inversion in the order of control from the material conditions to systems of expectations enables reflexive agents—human beings—to operate infra-reflexively across (vertical) levels and among (horizontal) compartments by changing their perspectives on the complexity (Latour, 1988; Pickering, 1995).

2. In the horizontal direction, we follow Parsons' (1968) proposal to consider the functional differentiation and symbolic generalization of the codes of communication as drivers of the increasing complexity in cultural evolution. Recalling another intuition of Herbert Simon (1973, at pp. 19 ff.), one can expect an alphabet of these codes; for example, power, love, truth, law, art, etc. Because of the various codes operating, the same communication can mean something quite different in terms of its affective value, its truth value, or how power is reproduced in communications (Parsons, 1963a and b; 1968; Künzler, 1987; Luhmann, 1974; 1997). The complexity is increased because the codes are also recombined in their instantiations (Hoffmeyer & Emmeche, 1991). A dynamics of differentiation among the codes versus integration in instantiations is thus to be specified. Based on decoding and recoding, for example, differently codified expectations about markets and technologies can be recombined into new technological options (Arthur, 2009; Cowan & Foray, 1997). A technological evolution can thus be generated as a retention mechanism of the cultural evolution of possible expectations (Dubois, 2003).

---

[1] In Giddens' structuration theory, these next-order structures are considered as rules and resources. From this perspective, structures exist only as memory traces, the organic basis of human knowledgeability, and as instantiated in action (Giddens, 1984, p. 177). Giddens (1979, pp.81 f.) formulates as follows: "The communication of meaning in interaction does not take place separately from the operation of relations of power, or outside the context of normative sanctions. (…) (P)ractices are situated within intersecting sets of rules and resources that ultimately express features of the totality."

**The generation of meaning from (Shannon-type) information**

How can the processing of meaning be conceptualized by elaborating on Shannon's theory despite the author's explicit statement that the "semantic aspects of communication are irrelevant to the engineering problem" (Shannon, 1948, at p. 3)? As a first step in the specification of the relevance of Shannon's engineering model for developing a theory of meaning, Weaver (1949, at p. 26) proposed two "minor additions" to Shannon's well-known diagram of a communication channel (Figure 1), as follows:

> "One can imagine, as an addition to the diagram, another box labeled "Semantic Receiver" interposed between the engineering receiver (which changes signals to messages) and the destination. This semantic receiver subjects the message to a second decoding, the demand on this one being that it must match the statistical semantic characteristics of the message to the statistical semantic capacities of the totality of receivers, or of that subset of receivers which constitute the audience one wishes to affect.
> Similarly one can imagine another box in the diagram which, inserted between the information source and the transmitter, would be labeled "semantic noise," the box previously labeled as simply "noise" now being labeled "engineering noise." From this source is imposed into the signal the perturbations or distortions of meaning which are not intended by the source but which inescapably affect the destination. And the problem of semantic decoding must take this semantic noise into account."

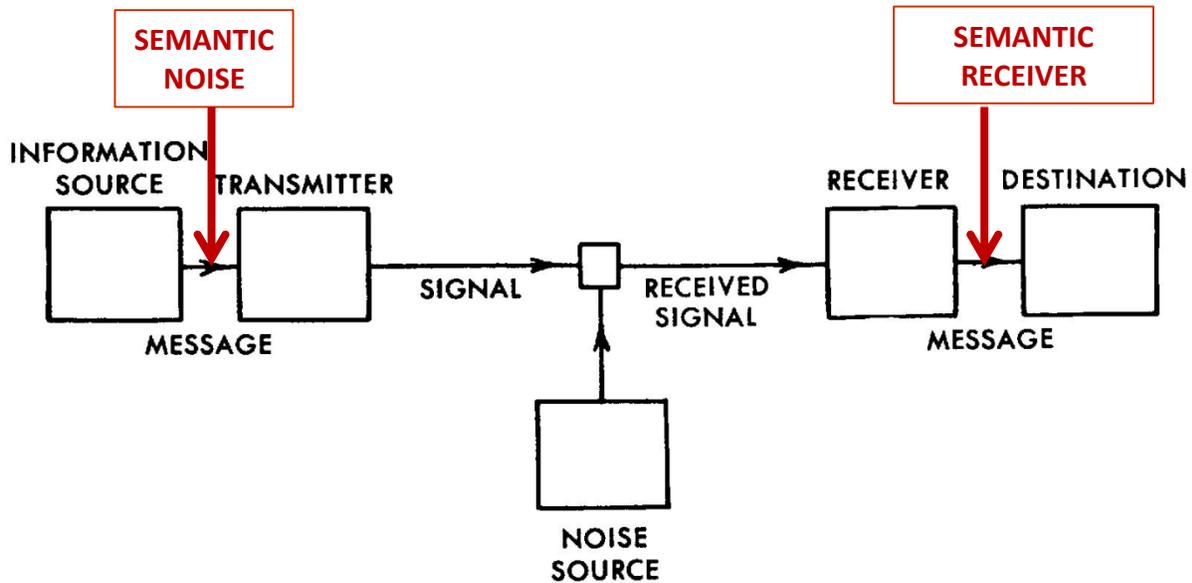

**Figure 1**: Weaver's (1949) "minor" additions penciled into Shannon's (1948) original diagram.

Since the "semantic receiver" recodes the information in the messages (received from the "engineering receiver" who only changes signals into messages) while having to assume the possibility of "semantic noise," a semantic relationship between the two new boxes can also be envisaged. Given Shannon's framework, however, this relation cannot be considered as another information transfer—since semantics are defined as *external* to Shannon's engineering model.

Semantics are not based on specific communications, but on relations among *patterns of relations* or, in other words, correlations. Two competing firms, for example, may have highly correlating patterns of relations with clients, but no relation with each other (Burt, 1982). The correlations among the distributions span a vector space in a topology different from the network space of relations (Appendix 1).[2] Two synonyms, for example, can have the same position (and meaning) in the vector space, yet never co-occur in a single sentence as a relation. Meanings can be shared also without a direct relation.

In the case of a single relation, the relational distance is not different from the correlational one; but in the case of relations involving three (or more) agents, the distances in the vector space are different from the Euclidean distances in the network space. Simmel (1902) already noted that the transition from a group of two to three is qualitative. In a triplet, the instantiation of one or the other relation can make a difference for the further development of the triadic system of relations.

A system of relations can be considered as a semantic domain (Maturana, 1978). In other words, the sender and receiver are related in the graph of Figure 1, while they are correlated in terms of not necessarily instantiated relations in the background. The structure of correlations provides a latent background that provides meaning to the information exchanges in relations.[3] The correlations are based on the same information, but the representation in the vector space is different from the graph in the network space of observable relations. In other words, meaning is not added to the information, but the same information is delineated differently and considered from a different perspective (including absent relations; i.e., zeros in the distribution). As against Shannon-type information which flows linearly from the sender to the receiver, one can expect meanings to loop, and thereby, to develop next-order dimensionalities (Krippendorff, 2009a and b).

---

[2] Each pattern of relations can be considered as a vector. When two vectors stand orthogonal in the vector space, the correlation is zero. See the Appendix for further explanation.
[3] "Structures exist paradigmatically, as an absent set of differences, temporarily 'present' only in their instantiations, in the constituting moments of social systems." (Giddens, 1979, p. 64).

**The third and fourth dimensions of the probabilistic entropy**

A matrix of communications is shaped when one adds a second dimension (of codes) to the single vector of communication in Figure 1. A matrix can also be considered as a two-dimensional probability distribution—different from the one-dimensional probability distribution in a vector. When the communication matrix—of information processing and meaning processing—is repeated over time, one obtains a three-dimensional array because time is added as a third dimension (Figure 2a). In such a three-dimensional array, the development of information can also be considered in terms of trajectories; the uncertainty is then organized historically (over time). A four-dimensional array or hyper-cube of information is more difficult to imagine or represent graphically: unlike a (three-dimensional) trajectory, a four-dimensional array can contain a next-order regime that feeds back on its historical development along trajectories (Dosi, 1982).

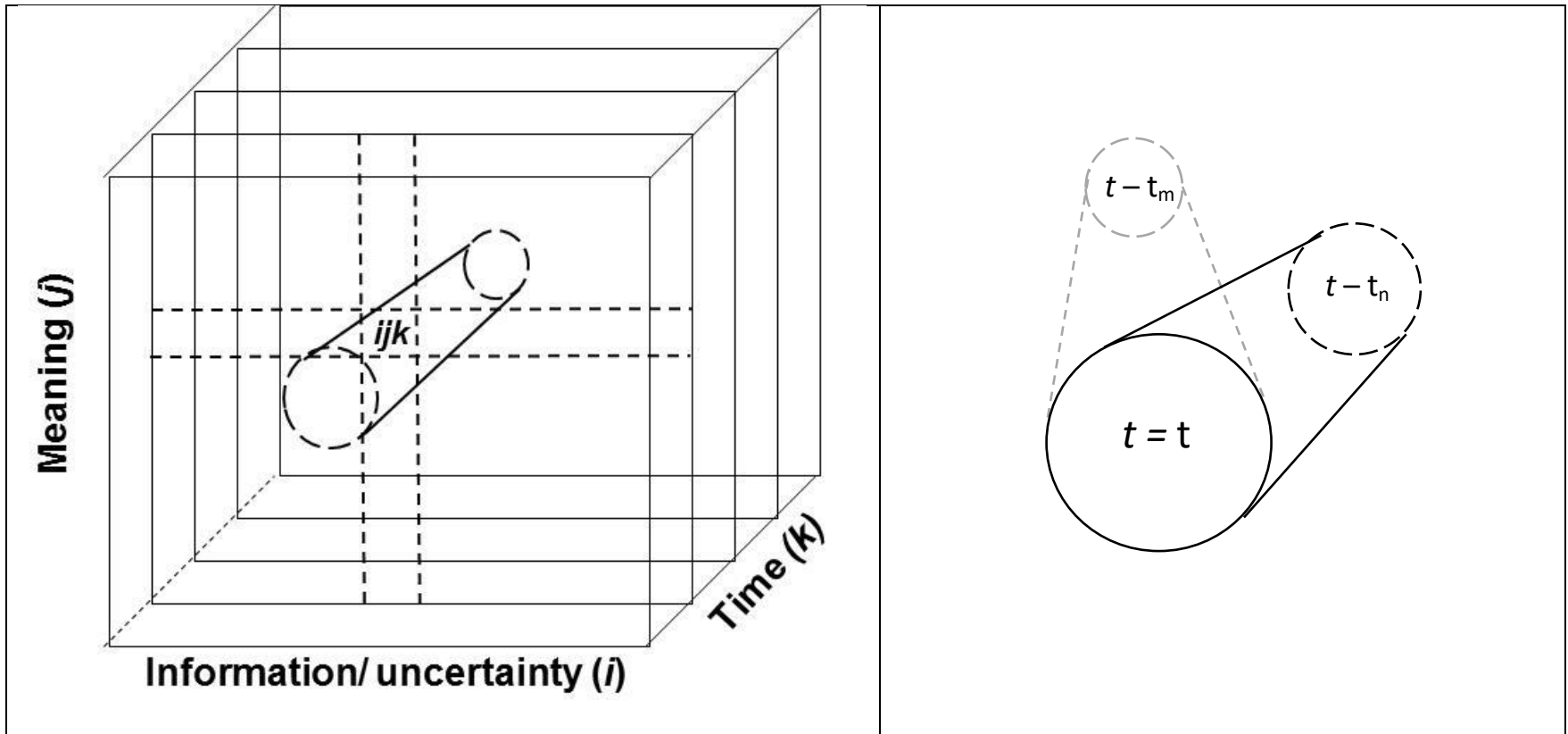

**Figure 2a (left) and 2b (right)**: A three-dimensional array of information can contain a trajectory; a four-dimensional hypercube contains one more degree of freedom and thus a variety of possible trajectories. Source: Leydesdorff (1996), p. 289.

One can consider the next-order regime as having one more degree of freedom that allows it to select among the possible trajectories in three dimensions as representations of its past (Figure 2b). This additional selection implies a reflection by the system. The reflection is performative in the present and can therefore be considered as the self-organization of an adaptive system. In the four-dimensional system, one representation of its history in three dimensions can be acknowledged (weighted) more than another.

We can consider ourselves as psychologies with the reflexive capacity to reconstruct the possible representations of our history. Luhmann (e.g., 1986a) suggested modeling the social system of communications as a system without psychological consciousness, but with an equal level of complexity. Whereas a psychological system is centered on the *individuum* and will therefore tend to integration (Haken & Portugali, 2014), the communication system is distributed (as a "*dividuum*" ; Luhmann, 1984: 625; cf. Nietzsche, [1878] 1967: 76) and has the option of exploiting the additional degree of freedom for differentiation. Different from a high culture, the modern society is based on prevailing differentiation among codes of communication, so that a set of juxtaposed coordination mechanisms can be used. The different coordination mechanisms are partially integrated when the systems are instantiated in terms of historical organization and action.

In terms of evolution theory, the coordination mechanisms (e.g., the market) can also be considered as selection mechanisms which operate with different criteria. The recursion of selection at the structural level leads to second-order variation: selections can be selected for stabilization (in history), and some stabilizations can be selected for globalization (at the regime level). For subsequent selections, however, the historical origin of a variation is not always relevant. Thus, the system of expectations may continuously loop into itself at different levels and from different perspectives.

**Levels B and C in the Shannon diagram**

Weaver (1949, p. 24) suggested taking Shannon's original diagram as a representation of "level A" which can be complemented with more levels (B and C) that represent how meaning is conveyed at level B, and how and why the received meaning can affect behavior (at level C)? Elaborating on Shannon's model and Weaver's addition as depicted in Figure 1, we propose Figure 3 as a scheme for levels B and C.

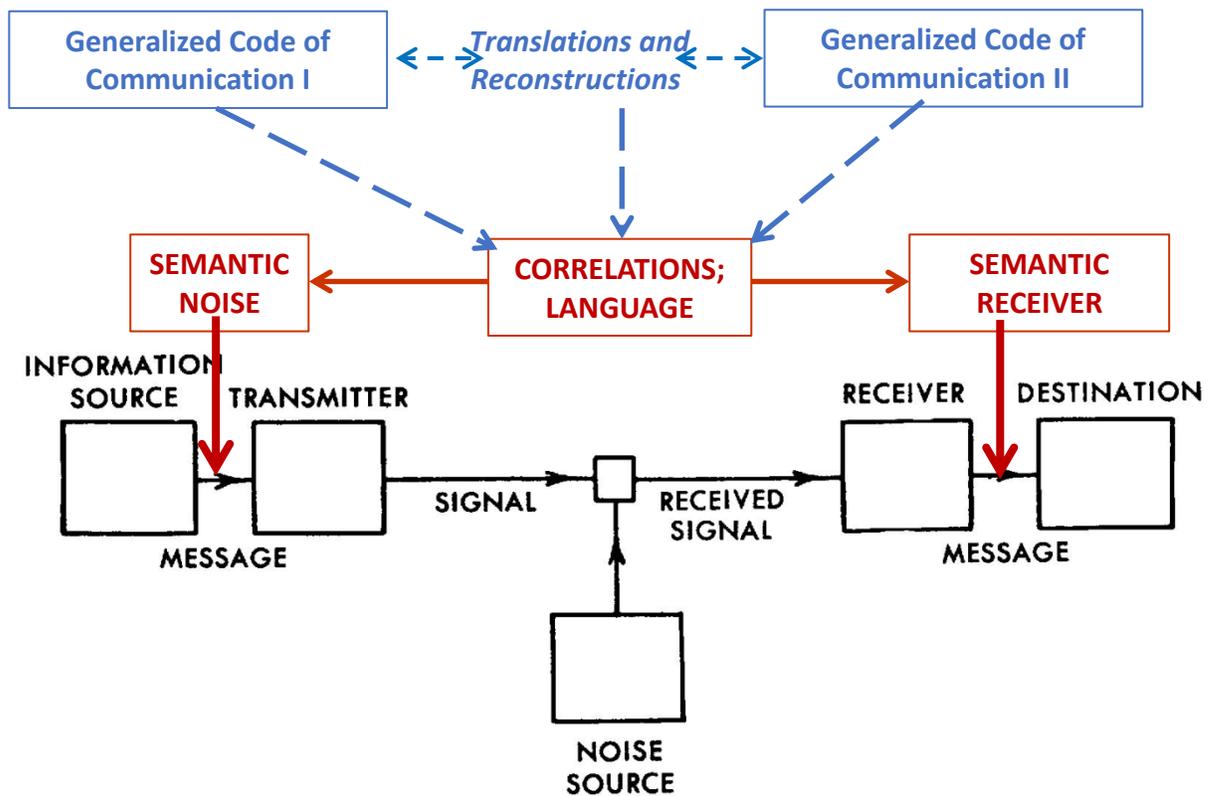

**Figure 3**: Levels B and C added to the Shannon diagram (in red-brown and dark-blue, respectively).

We specified above that the relation between the semantic receiver and semantic noise is based on correlations among sets of relations at level A. In the vector space (level B), meanings can be shared, but not communicated (because otherwise one operates at level A). The use of language facilitates, supports, and potentially reinforces the options for sharing meaning. Natural languages (at level B), however, can be considered as the as yet undifferentiated and therefore common medium of communication. Codes of communication are used at the symbolic level C for regulating the use of language. The codes enable us, among other things,[7] to short cut the communication; for example, by paying the market price of a good instead of negotiating this price using language. In our opinion, the codes of communication are thus candidates for Weaver's level C: the codes and their combinations enable us to make the communications far more precise and efficient than is possible in natural languages.

Talcott Parsons (1968, p. 440) provided a sociological appreciation of the operations at level C as follows:

> "At the cultural level [language] is clearly the fundamental matrix of the whole system of media. Large-scale social systems, however, contain more specialized media (if you will, specialized "languages"), such as money, power, and influence (see Parsons 1963a; 1963b). Such media, like language, *control* behavior in the processes of interaction."

In addition to symbolic, Parsons (1963a and b; 1968) characterized these media as "generalized" with a reference to Mead's (19324: pp. 154 ff.) "generalized other." Luhmann (1974) further argued that "symbolically generalized media of communication" would have to be binary—like true and false in logics—in order to be binding (Luhmann, 1984: 316f.; 1995: 233f.; cf. Künzler, 1987: 329 ff.). In our opinion, they can be more complex than one-dimensional and binary (cf. Hoffmeyer & Emmeche, 1991). For example, Herbert Simon (1973b) argued in favor of truth-finding and puzzle-solving as combined "logics" in scientific discovery. However, this operationalization has also remained chiefly a philosophical appreciation of the evolutionary process (cf. Popper, 1959 [1935]) more than a proposal for empirical operationalization (cf. Newell & Simon, 1972).

---

[7] Spelling rules, syntax, and pragmatics can also be considered as codes in the use of language, but we focus on the semantics.

In our opinion, the sciences evolve as systems of expectations rationalized by arguments in discourses; after a further development, the criteria may also have changed (Fujigaki, 1998; Kuhn, 1962). Using the conceptualization of the codes as the latent dimensions (principal components or "eigenvectors";[8] von Foerster, 1960; cf. von Glasersfeld, 2008, at p. 64, 4n.) of the communication matrix, one can appreciate the uncertain and evolving character of these codes of communications. From this perspective, the designation of these structures remains a historical appreciation.

For example, in the case of the science system, Luhmann (1990) argued for true/not-true as the code that provides a binary criterion for quality control in scholarly discourse. However, in the empirical sciences, the truth of statements is not unambiguous: some statements can be more true or less false than others. Since the sciences develop as discursive knowledge, uncertainty is always present. In a study of the debates about oxidative phosphorylation—which led to the Nobel Price of Chemistry for Peter Mitchell in 1978—Gilbert & Mulkay (1984) found that statements were relabeled using different repertoires when they were considered true or erroneous from the perspective of hindsight.

---

[8] Using linear algebra: for any function f, if a and $\lambda$ exist such that $f(a) = \lambda f(a)$, then a is called the eigenvector and $\lambda$ the eigenvalue (Achterbergh & Vriens, 2009, p. 84n.). In the case of a communication matrix W, $WA = \lambda A$ with $A$ as eigenvector and $\lambda$ as eigenvalue. Eigenvectors can also be considered as pointing to densities consequential to the recursive operations of self-organizing systems (von Foerster, 1960, 1982).

Codes can also be nested. For example, different specialties may operate with different codes while sharing some general criteria for the quality control of scholarly communications. Similarly, in economic transactions, a variety of payment methods can be distinguished under the umbrella of an economic logic that differs from a scholarly or normative one (Boudon, 1979; Bourdieu, 1976).

In his last book, Pierre Bourdieu (2004, at p. 83) precisely formulated a reflection on the empirical study of the sciences, as follows:

> Each field (discipline) is the site of a specific legality (a nomos), a product of history, which is embodied in the objective regularities of the functioning of the field and, more precisely, in the mechanisms governing the circulation of information, in the logic of the allocation of rewards, etc., and in the scientific habitus produced by the field, which are the condition of the functioning of the field. […]
>
> What are called epistemic criteria are the formalization of the 'rules of the game' that have to be observed in the field, that is, of the sociological rules of interactions within the field, in particular, rules of argumentation or norms of communication. Argumentation is a collective process performed before an audience and subject to rules.

Bourdieu (at p. 78) calls this a "Kantian"—that is, transcendental—transition from "objectivity" to "intersubjectivity" as the carrying ground of scientific inferences. However, the philosopher most associated with this transition is Edmund Husserl, who criticized the increasingly empiristic self-understanding of the modern (European) sciences (Husserl, [1935/36] 1962). According to Husserl ([1929] 1960, at p. 155), the possibility to communicate expectations intersubjectively grounds the empirical sciences "in a concrete *theory of science"*. One tests expectations entertained at the supra-individual level (in discourses) against observations, and the observations can update the expectations since they can function as arguments.

We shall take from Husserl that the self-organizing codes of the communication at level C are not material, but belong to our reality as structures of expectations or *res cogitans.* Popper (1972) denoted this domain as World 3, but neither he nor Husserl specified the evolutionary dynamics of expectations in terms of communications (Luhmann, 1986b). We submit that *res cogitans* can be expected to develop in terms of redundancies instead of probabilistic entropy, unlike the material world (*res extensa*) where the Second Law prevails. Language and the symbolic media of communication enable us to multiply meanings as options at a speed much faster than can historically be realized. The codes of communications provide us with horizons of other possible meanings. The different codes can be recombined and reconstructed in translations among differently coded meanings. We shall argue that at level B, meanings are instantiated in specific combinations of codes, while at level C the codes themselves evolve in response to the historical integrations in the instantiations.

**The transformation of hitherto "impossible" options into technological feasible ones**

Whereas historical developments unfold with the arrow of time—and are necessarily related to the generation of entropy—expectations enable us to use possible future states in the present, i.e., against the arrow of time. The dynamics of expectations therefore are very different from organizational dynamics. Under specifiable conditions, the interactions among differently coded expectations can generate redundancy (that is, negative entropy). Redundancy enriches a system with new options that are available for realization.

The redundancy $R$ is defined in information theory as the fraction of the capacity of a communication channel that is not used. In formula format:

$$R = 1 - \frac{H}{H_{max}}$$
$$= \frac{H_{max} - H}{H_{max}} \tag{1}$$

As is well-known, Shannon's (1948) probabilistic entropy ($H$) is coupled to Gibbs' formula for thermodynamic entropy $S = k_B * H$. In Gibbs' equation, $k_B$ is the Boltzmann constant that provides the dimensionality Joule/Kelvin to $S$, while $H$ is dimensionless and can be measured in bits of information (or any other base of the logarithm) given a probability distribution (containing uncertainty). The Second Law of thermodynamics states that entropy increases with each operation, and Shannon-type information is accordingly always positive.

Brooks & Wiley (1986) noted that in the case of an evolving (e.g., biological) system, not only the observed (probabilistic) entropy ($H_{obs}$) of the system increases, but also the maximum entropy ($H_{max} = \ln N$)—that is the system's capacity— increases as the total number of possible options (states) $N$ also increases. The difference between the maximum entropy and the realized entropy is provided by the options that are available, but have not yet been used. From the perspective of information theory, these surplus options are redundant. Using Brooks & Wiley's (1986, at p. 43) illustration in Figure 4a, we added green to the redundancy as part of the evolving entropy. Redundancy provides a measure of the options that were not realized by the system, but could have been realized. Kauffman (2000), for example, calls these possible realizations "adjacent."

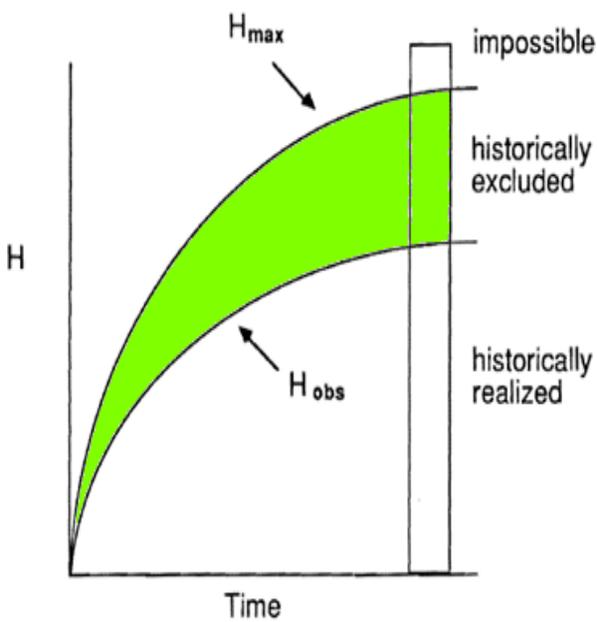 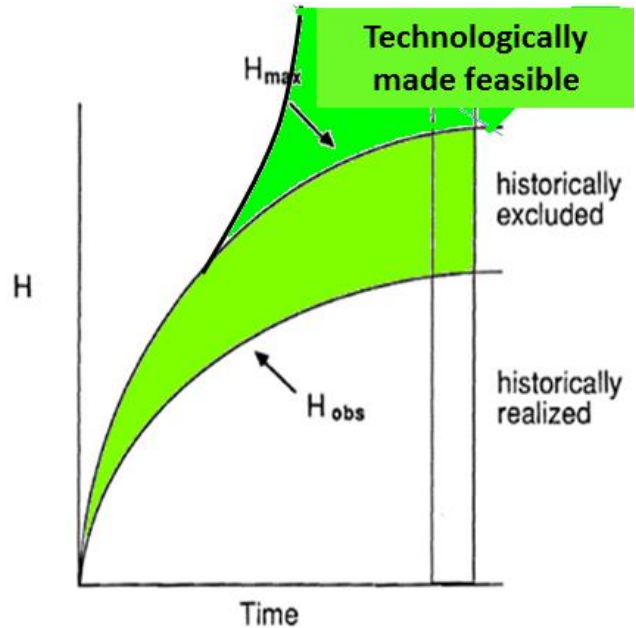

**Figure 4a**: The development of entropy ($H_{obs}$), maximum entropy ($H_{max}$), and redundancy ($H_{max} - H_{obs}$). Source: Brooks & Wiley (1986, at p. 43).

**Figure 4b:** Hitherto impossible options are made possible because of cultural and technological evolution.

Above the green area, Brooks & Wiley (1986, at p. 43) added the label "impossible" as a legend (see Figure 4a). In Figure 4b, we have added the domain "technologically made feasible" to this latter area in order to introduce how the generation of new options (and hence increased redundancy) can be enhanced by a model of cultural evolution which includes the levels B and C. An intentional system operates by adding new options (redundancy) without necessarily realizing them (as probabilistic entropy).

The codes regulate the generation of redundancies at interfaces from above, whereas Shannon entropy is continuously generated in the historical process from below. The latter process is linear, whereas expectations can circulate before being organized in realizations. Redundancy is generated when two (or more) perspectives on the same information are operating at an interface, as in the case of introducing a new technology in a market or when writing a report based on scholarly arguments for a government agency. In such cases, one needs text that can be read using the various codes involved (Fujigaki & Leydesdorff, 2000). The redundancy observable in the green surfaces of Figure 4b is generated by the recombination of sufficiently different expectations. Let us try to specify this process in information-theoretical terms.

*a. Mutual redundancy between two differently coded systems*

In Figure 5, the overlap between uncertainties in two variables $x_1$ and $x_2$ is depicted as two sets. The mutual information or transmission ($T_{12}$) is then defined as follows:

$$T_{12} = H_1 + H_2 - H_{12} \tag{2}$$

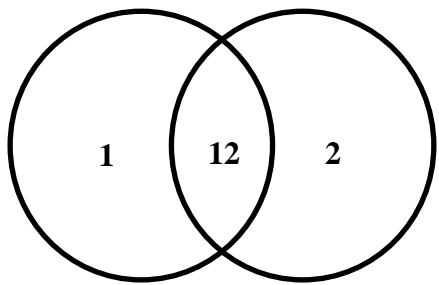

**Figure 5**: Overlapping uncertainties in two variables $x_1$ and $x_2$.

Note that the addition of entropies accords with the rules of set theory. Alternatively, one can consider the overlap as a redundancy: the same information is appreciated twice. In addition to $H_1$ and $H_2$, the overlap contains redundancy as a surplus of information, as follows:

$$Y_{12} = H_1 + H_2 + T_{12} = H_{12} + 2T_{12} \tag{3}$$

The mutual redundancy $R_{12}$ at the interface between the two sets can now be found by using $Y_{12}$ instead of $H_{12}$ in Eq. 2, as follows:

$$\begin{aligned} R_{12} &= H_1 + H_2 - Y_{12} \\ &= H_1 + H_2 - (H_{12} + 2T_{12}) \\ &= H_1 + H_2 - ([H_1 + H_2 - T_{12}] + 2T_{12}) \\ &= -T_{12} \end{aligned} \tag{4}$$

Since $T_{12}$ is necessarily positive (Shannon, 1948, p. 53), it follows from Eq. 4 that $R_{12}$ is always negative and therefore by definition a redundancy. This reduction of the uncertainty can be measured in bits of information with a negative sign. In other words, this redundancy cannot be a Shannon-type information, since the latter information is necessarily positive (Krippendorff, 2009a). Using mutual redundancy, one no longer measures a historical process—generating uncertainty—but a process in the realm of expectations: future states are represented in the present (Dubois, 1998).

*b. Redundancy in three and four dimensions*

For the three-dimensional case and using Figure 6, one can define, in addition to the two-dimensional values of $Y$ (in Eq. 3), a three-dimensional value including the redundancies (in the respective overlaps) as follows:

$$Y_{123} = H_1 + H_2 + H_3 + T_{12} + T_{13} + T_{23} + T_{123} \tag{5}$$

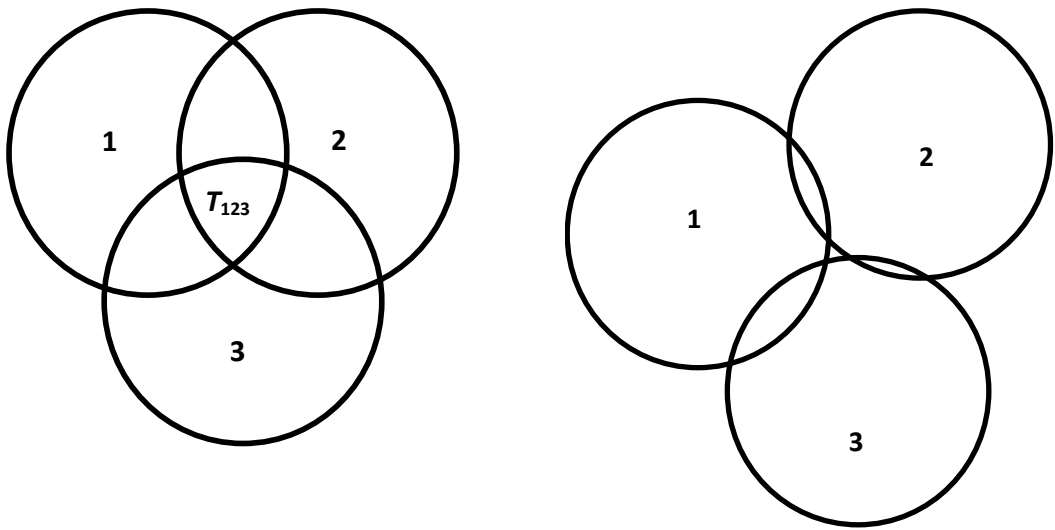

**Figure 6**: Overlapping uncertainties in three variables $x_1$, $x_2$, and $x_3$: two configurations with opposite sign of $T_{123}$.

Using information theory, however, one would make the following corrections for double counting in the overlaps (by subtracting; see Figure 6):

$$H_{123} = H_1 + H_2 + H_3 - T_{12} - T_{13} - T_{23} + T_{123} \qquad (6)$$

It follows that the difference between the uncertainty in the historical system ($H_{123}$ in Eq. 6) and the system of expectations ($Y_{123}$ in Eq. 5) is:

$$Y_{123} - H_{123} = +2T_{12} + 2T_{13} + 2T_{23}$$

$$Y_{123} = H_{123} + 2T_{12} + 2T_{13} + 2T_{23} \qquad (7)$$

Furthermore, the mutual information in three dimensions can be derived in information theory using the Shannon formulas (e.g., Abramson, 1963, at p. 129; McGill, 1954; Yeung, 2008) as:

$$T_{123} = H_1 + H_2 + H_3 - H_{12} - H_{13} - H_{23} + H_{123} \qquad (8)$$

Using *Y*-values instead of *H*-values for the joint entropies in Eq. 8, one obtains the mutual redundancy in three dimensions as follows:

$$\begin{aligned} R_{123} &= H_1 + H_2 + H_3 - (H_{12} + 2T_{12}) - (H_{13} + 2T_{13}) - (H_{23} + 2T_{23}) \\ &\quad + (H_{123} + 2T_{12} + 2T_{13} + 2T_{23}) \\ &= T_{123} \end{aligned} \qquad (9)$$

In the three-dimensional case, the mutual redundancy is thus equal to the mutual information in three dimensions. Furthermore, Leydesdorff & Ivanova (2014, at p. 392) show that in the case of four dimensions $R_{1234} = -T_{1234}$. The sign of the mutual redundancy alternates with the number of dimensions. This corrects for the otherwise inexplicable sign changes in the mutual information with increasing dimensionality. This sign change of the mutual information with dimensions is a well-known problem in information theory, but beyond the scope of the present study.[9] In other words, mutual redundancy is a consistent measure of negative entropy, while mutual information is not, because of its sign changes with the dimensionality. We prove this claim in the next section by generalizing the formulation.

*c. Generalization*

Eq. 8 can be rewritten as follows:

$$T_{123} = H_1 + H_2 + H_3 - H_{12} - H_{13} - H_{23} + H_{123} \tag{8}$$

$$T_{123} = [(H_1 + H_2 - H_{12}) + (H_1 + H_3 - H_{13}) + (H_2 + H_3 - H_{23})] + [H_{123} - H_1 - H_2 - H_3]$$

$$T_{123} = [T_{12} + T_{13} + T_{23}] + [H_{123} - H_1 - H_2 - H_3] \tag{10}$$

The second bracket in Eq. 10 makes a negative contribution, because of the subadditivity of the entropy: $H(x_1, \ldots, x_n) \leq \sum_1^n H(x_i)$, which holds for any dimension $n \geq 2$. The terms in the first bracket of Eq. 10 are strictly positive. As noted, the sign of the resulting value of $T_{123}$ depends on the empirical configuration (as indicated by the two configurations in Figure 6).

---

[9] Krippendorff (2009b, at p. 670) provided a general *notation* for this alteration with changing dimensionality—but with the opposite sign (which further complicate the issue; cf. Leydesdorff, 2010: 68)—as follows:

$$Q(\Gamma) = \sum_{X \subseteq \Gamma} (-1)^{1+|\Gamma|-|X|} H(X) \tag{9}$$

In this equation, $\Gamma$ is the set of variables of which $X$ is a subset, and $H(X)$ is the uncertainty of the distribution; $|\Gamma|$ is the cardinality of $\Gamma$, and $|X|$ the cardinality of $X$.

It follows (inductively) that for any given dimension $n$, one can formulate combinations of mutual informations corresponding to $\sum_1^n H(x_i) - H(x_1, \dots, x_n)$ that are by definition positive (or zero in the null case of complete independence). For example (up to four dimensions) as follows:

$$0 \leq \sum_{i=1}^{n=2} H(x_i) - H(x_1, x_2) = T_{12}$$

$$0 \leq \sum_{i=1}^{n=3} H(x_i) - H(x_1, x_2, x_3) = \sum_{ij}^{3} T_{ij} - T_{123} \quad (11)$$

$$0 \leq \sum_{i=1}^{n=4} H(x_i) - H(x_1, x_2, x_3, x_4) = \sum_{ij}^{6} T_{ij} - \sum_{ijk}^{4} T_{ijk} + T_{1234}$$

where the sums on the right-hand side are over the $\binom{n}{k}$ permutations of the indices. This relation can be extended for general $n$ as,

$$0 \leq \sum_{i=1}^{n} H(x_i) - H(x_1, \dots, x_n)$$

$$= \sum_{ij}^{\binom{n}{2}} T_{ij} - \sum_{ijk}^{\binom{n}{3}} T_{ijk} + \sum_{ijkl}^{\binom{n}{4}} T_{ijkl} - \cdots + (-1)^{1+n} \sum_{ijkl\dots(n-1)}^{\binom{n}{n-1}} T_{ijkl\dots(n-1)} +$$

$$(-1)^n \sum_{ijkl\dots(n)}^{\binom{n}{n}} T_{ijkl\dots(n)} \quad (12)$$

where the last term on the right-hand side is equal to $(-1)^n T_{1234\dots n}$. Returning to the relation between $R_{12}$ and $T_{12}$, it now follows instructively that:

$$R_{12} = -T_{12}$$

$$= H(x_1, x_2) - \sum_1^2 H(x_i) \leq 0 \quad (13)$$

and the analogous relations for $R_{123}$ and $R_{1234}$ follow in the same way from Eq. (12). More generally, in the case of more than two dimension, $n > 2$:

$$R_n = (-1)^{1+n} T_{1234\ldots n} = [H(x_1, \ldots, x_n) - \sum_1^n H(x_i)]$$

$$+[\sum_{ij}^{\binom{n}{2}} T_{ij} - \sum_{ijk}^{\binom{n}{3}} T_{ijk} + \sum_{ijkl}^{\binom{n}{4}} T_{ijkl} - \cdots + (-1)^{1+n} \sum_{ijkl\ldots(n-1)}^{\binom{n}{n-1}} T_{ijkl\ldots(n-1)}] \quad (14)$$

The left-bracketed term of Eq. 14 is necessarily negative entropy (because of the subadditivity of the entropy), while the configuration of the remaining mutual information relations contribute a second term on the right which is positive (see the set of Equations 11 above). In other words, we model here the generation of redundancy on the one side versus the historical process of uncertainty generation in relating on the other, as an empirical balance in a system that operates with more than two codes (e.g., alphabets; Abramson, 1963, pp. 127 ff.). When the resulting $R$ is negative, self-organization prevails over organization in the configuration under study, whereas a positive $R$ indicates conversely a predominance of organization over self-organization as two different subdynamics.

*d. Clockwise and anti-clockwise rotations*

As soon as the relation between two subdynamics (or agents) is extended with a third, the third may feedback or feed-forward on the communication relation between the two, and thus a system is shaped (Sun & Negishi, 2010). This principle is known in social network analysis as "triadic closure" (Bianconi *et al.*, 2014; De Nooy & Leydesdorff, 2015). Triadic closure can be considered as the basic mechanism of systems formation. When three selection environments operate on variation in the interactions among them, the communication can proliferate auto-catalytically using each third mechanism as a feedback on or feed-forward to bi-lateral relations. At a next moment, the cycling may take control as a vortex (Ivanova & Leydesdorff, 2014b).

Ulanowicz (2009 at p. 1888) depicted this possibility of auto-catalysis as follows:

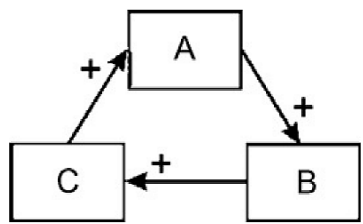

**Figure 7**: Schematic of a hypothetical three-component autocatalytic cycle.
(Source: Ulanowicz, 2009, at p. 1888, Figure 3.)

A second cycle with the reverse order of the operations is equally possible; the two cycles can be modeled as two vectors with three dimensions (A, B, and C), and this system can then be simulated in terms of rotations of the vectors. One vector can be understood as corresponding to the tendency of historical realization and the other to self-organization. Using simulations, Ivanova & Leydesdorff (2014a) showed that the operation of these two three-dimensional vectors upon each other generates a value of $R$ depending on the configuration. As we showed above, $R$ can also be measured.

The one sign of $R$ can be associated with clockwise and the other with anti-clockwise rotation, whereas the values of the two terms in Eq. 14 measure the relative weights of the two rotations. In other words, mutual redundancy indicates the size of the footprint of the self-organizing fluxes of communication on the historical organization in the instantiation. The cycles can be vicious or virtuous in the sense of providing opportunities or exploiting existing ones.

**Summary and conclusions**

We have extended Shannon's model of communication (at level A) with two levels (B and C) that change the linear model into an evolutionary one because feedback and feed-forward loops are possible among the levels. At level A, information is transmitted; at level B, information is organized and thus made meaningful in a vector-space. Reflexivity reveals that this vector space is constructed and therefore a potential subject of reconstruction: the possibility of reconstruction opens horizons of meaning (level C). These horizons can be expected to evolve along the eigenvectors of the communication matrix in different directions. Whereas the common language at level B tends to integration (into organization), horizontal differentiation among the codes at level C increases the communication capacity of the system.

Codes of communication are no longer actor-attributes, but operate on the communications among human beings reflexively—that is, by co-constructing the meaning of the communication (Luhmann, 1984). In other theoretical contexts, one can also consider the codes as virtual coordination mechanisms (e.g., Giddens' structuration theory) or as selection mechanisms (e.g., Nelson & Winter, 1982; Dosi, 1982). As Andersen (1992, p. 14) noted specification of "What is evolving?" becomes a relevant question when selection is no longer given by nature (cf. Boulding, 1978, p. 33). The question arises, under which conditions can the different selection mechanisms be expected to co-evolve and lead to new options for realizing innovations?

We showed that redundancies can be generated at interfaces among systems as sets of relations which are structured by codes, whereas in historical relations (instantiations) only variety is generated. Biological evolution theory assumes *variation* as a driver and *selection* to be naturally given, while cultural evolution is driven by individuals and groups who make conscious decisions on the basis of potentially different criteria (Newell & Simon, 1972; Petersen *et al*., 2016).

Note that the interactions among the codes in the instantiations generate the redundancies. In Luhmann's theory, these interactions are held to be impossible: the binary codes close the autopoiesis of systems and subsystems operationally. Whereas biological systems gain in complexity by closing themselves operationally, systems of expectations can disturb one another "infra-reflexively" (Latour, 1988, at pp. 169 ff.). Cultural evolution can therefore be much faster than biological evolution. The "avoidance of redundancy" is an objective in Luhmann's model of autopoiesis and functional differentiation (e.g., 2013, p. 98), whereas redundancy generation is considered crucial for the advancement of society in our model. Redundancy provides new options for technological development that are not yet realized, but can be envisaged.

In addition to vertical differentiation, the assumption of horizontal differentiation is needed for understanding the evolving complexity. The differentiation and the ensuing pluriformity of the competing coordination mechanisms break the structural formations of autopoietic systems so that the established meanings can be interrupted by other possible meanings. The generation of redundancy can thus enter the historical instantiations, and under the condition of self-reinforcing loops, can thereby tip the balance towards the prevalence of evolutionary self-organization over historical organization. We have shown how the trade-off between historical organization and self-organization over time can be traced by the measurement of mutual redundancies.

When three or more selection mechanisms operate, auto-catalysis is an option, and options can then be generated at an increasing pace. Thus, horizontal differentiation is a necessary component of self-organization in the vertical dimension. The warp and the woof of meaning generation and self-organization are not harmoniously integrated as in textiles, but differentiated and disturbing one another. The layers are not hierarchical, but operating in parallel. These horizontal and vertical dynamics lead to a fractal manifold in different directions. Through breakages (and hence puzzles) new options can be generated.

The generation of redundancy proceeds in a domain of expectations about options that do not (yet) exist, but that one can reflexively entertain. By turning away from an objectivistic self-understanding of the sciences, we find room for a general theory of meaning and knowledge-generation that we have depicted as an extension of Shannon's theory (Figure 3). Whereas Shannon felt the need to explicitly deny this extension, Weaver understood it as the theory's proper intension.

**Acknowledgement**
II acknowledges support of the Basic Research Program at the National Research University Higher School of Economics (HSE) and the Global Competitiveness Program of the Government of the Russian Federation.

**Appendix 1**

As an example of a set of relations which can be formalized within a coordinate system (vector space) defined by relations, consider two agents or firms A and B with similar relations to agents/clients [C1, C2, C3], but no relations between them. In the case of a Triple-Helix configuration, C1, C2, and C3 can be a university, another firm, and the government, respectively. A and B can be represented as two vectors constructed as follows:

|        | A | B | C1 | C2 | C3 |
|--------|---|---|----|----|----|
| Firm A | 1 | 0 | 1  | 0  | 1  |
| Firm B | 0 | 1 | 1  | 0  | 1  |

The Pearson correlation between these two distributions is $r = 0.167$ (*n.s.*). A natural measure of similarity between vectors is the cosine of the angle between them; in this case, $cos(A,B) = 0.667$. Figure A1 represents the relations in a three-dimensional Euclidean space of, for example, University-Industry-Government relations.

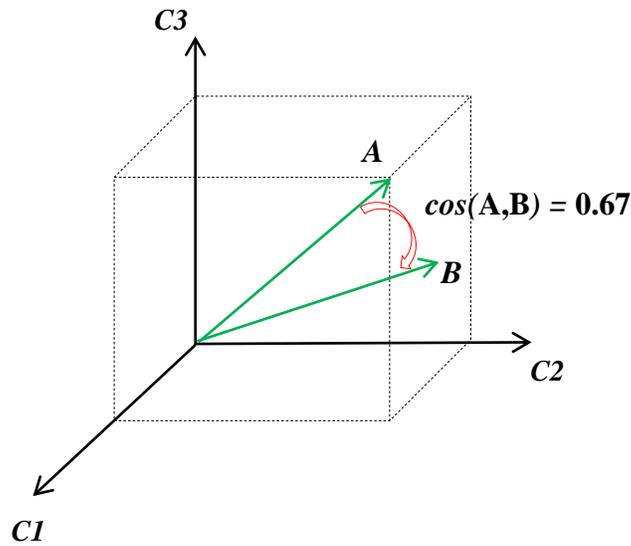

**Figure A1:** The two distributions as vectors in a three-dimensional ("Triple Helix") space. Source: Ivanova & Leydesdorff, 2014, p. 930.

Using the vectors themselves as non-orthogonal axes, one can span a vector space (Salton & McGill, 1983, pp. 120 ff.; Figure A2).

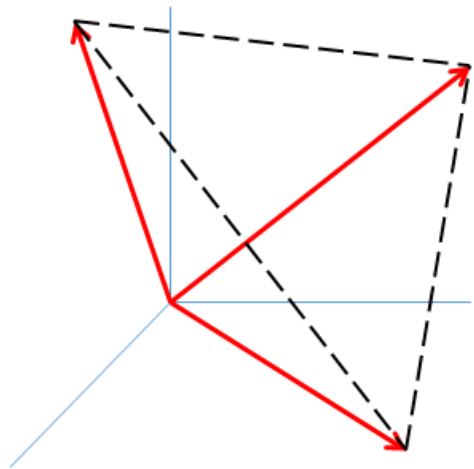

**Figure A2**: Vector space in the case of three vectors.

The vector space can be multi-dimensional; the dimensionality is determined by the number of vectors. Eigenvectors can be considered as vectors pointing to the centroids of clusters in a space reducing the dimensionality of the vector space (using, e.g., factor analysis or multi-dimensional scaling; Schiffman *et al*., 1981). See also footnote 8 above.